\documentclass[preprint]{revtex4}
\usepackage{graphicx}
\usepackage{amsmath}

\begin{document}

\title{Systematic Extraction of QGP Properties\footnote{Presented by S.~A.~Bass at "Critical Point and Onset of Deconfinement (CPOD) 2016", Wroc\l{}aw, Poland, May 30th - June 4th, 2016}}
\author{Jussi Auvinen}
\email{jaa49@phy.duke.edu}
\author{Jonah E. Bernhard}
\author{Steffen A. Bass}
\address{Department of Physics, Duke University, Durham, NC 27708, USA}

\begin{abstract}
We investigate the collision energy dependence of shear viscosity over the entropy density ratio $\eta/s$
in Au+Au collisions at $\sqrt{s_{NN}}=19.6, 39$, and $62.4$ GeV,
using Bayesian statistical analysis and Gaussian process emulators
to explore the full input parameter space of a transport+hydrodynamics hybrid model.
The ratio is found to decrease as a function of collision energy,
supporting the results from previous studies performed with the same hybrid model.
\end{abstract}

\maketitle

\section{Introduction}
\label{sec:intro}

Treatment of the quark-gluon plasma evolution and hadronic freeze-out in relativistic heavy ion collisions is well established and largely understood.
Major success has been achieved in extracting QGP properties such as $\eta$/s \cite{Niemi:2015qia,Denicol:2015nhu,Bernhard:2016tnd,Karpenko:2015xea}.
Quantifying the uncertainties in the extracted QGP properties presents a major challenge, however.
Beyond the temperature dependence of the transport coefficients,
the effects of a possible critical point and finite $\mu_B$ need to be quantified.
In addition, the physics of the initial state and pre-equilibrium dynamics are still conceptually challenging 
and are a major source of uncertainty for the extraction of QGP properties.

Rigorous model to data comparisons are difficult to perform,
as there is a large number of interconnected parameters with non-factorizable data dependencies.
Experimental data have also correlated uncertainties.
These issues necessitate the utilization of novel optimization techniques, such as Bayesian statistics
and Markov Chain Monte Carlo (MCMC) methods.
As the simulations of heavy ion collisions require considerable amount of computational resources,
new interpolation techniques based on emulators are also needed to predict the model output.

These statistical methods have already been applied with great success to Pb+Pb collisions at the LHC \cite{Bernhard:2016tnd}.
In the following we extend the analysis to the Au+Au collisions in the RHIC beam energy scan.

\section{Hybrid model}
\label{sec:hybrid}

We simulate the heavy ion collisions at RHIC-BES energies using the hybrid model described in Ref.~\cite{Karpenko:2015xea}.
In this model, the initial state produced by the UrQMD hadron+strings cascade \cite{Bass:1998ca,Bleicher:1999xi}.
The earliest possible starting time for hydrodynamical evolution is when the two colliding nuclei have passed through each other: $\tau_0 \geq 2R_{\text{nucleus}}/\sqrt{\gamma_{CM}^2-1}$.
At the transport-to-hydro transition, the microscopic particle properties (energy, baryon number) are mapped to densities
using 3-D Gaussians with ``smearing'' parameters $R_{\text{trans}}$, $R_{\text{long}}$ (=$\sqrt{2}$ times Gaussian width $\sigma$).

The hydrodynamic evolution is done with (3+1)D relativistic viscous hydrodynamics \cite{Karpenko:2013wva},
with a constant value of $\eta/s$ throughout the evolution, which is provided as input.
Transition from hydro back to transport (``particlization'') is performed when the energy density $\epsilon$ in the hydro cells reaches the switching value $\epsilon_{SW}$.
The iso-energy density hypersurface is constructed using the Cornelius routine \cite{Huovinen:2012is}.

\section{Statistical analysis}
\label{sec:stats}

The basics of the Bayesian analysis procedure and model emulation have been described in detail in \cite{Novak:2013bqa,Bernhard:2015hxa}.

The Bayesian posterior probability distribution is sampled using the Markov chain Monte Carlo (MCMC) method,
which is a random walk in parameter space, where each step is accepted or rejected based on a relative likelihood.
We initialize $\mathcal{O}(1000)$ random walkers at random positions in the input parameter space.
As in \cite{Bernhard:2015hxa}, the input parameter combinations were sampled using the maximin Latin hypercube method,
which attempts to optimize the sample by maximizing the minimum distance between the design points.

To calculate the likelihood, one must be able to determine the model output $\vec{y}$ for an arbitrary input parameter combination $\vec{x}$.
As the simulations typically take several hours to run, it is highly impractical to run the full model during statistical analysis.
Instead, the model is emulated with Gaussian processes, which provide very general, non-parametric interpolation of the physics model,
where the uncertainty related to the given estimate is included in a natural way. 

\section{Results}
\label{sec:results}

\begin{figure}
\centering
\includegraphics[width=6.2cm]{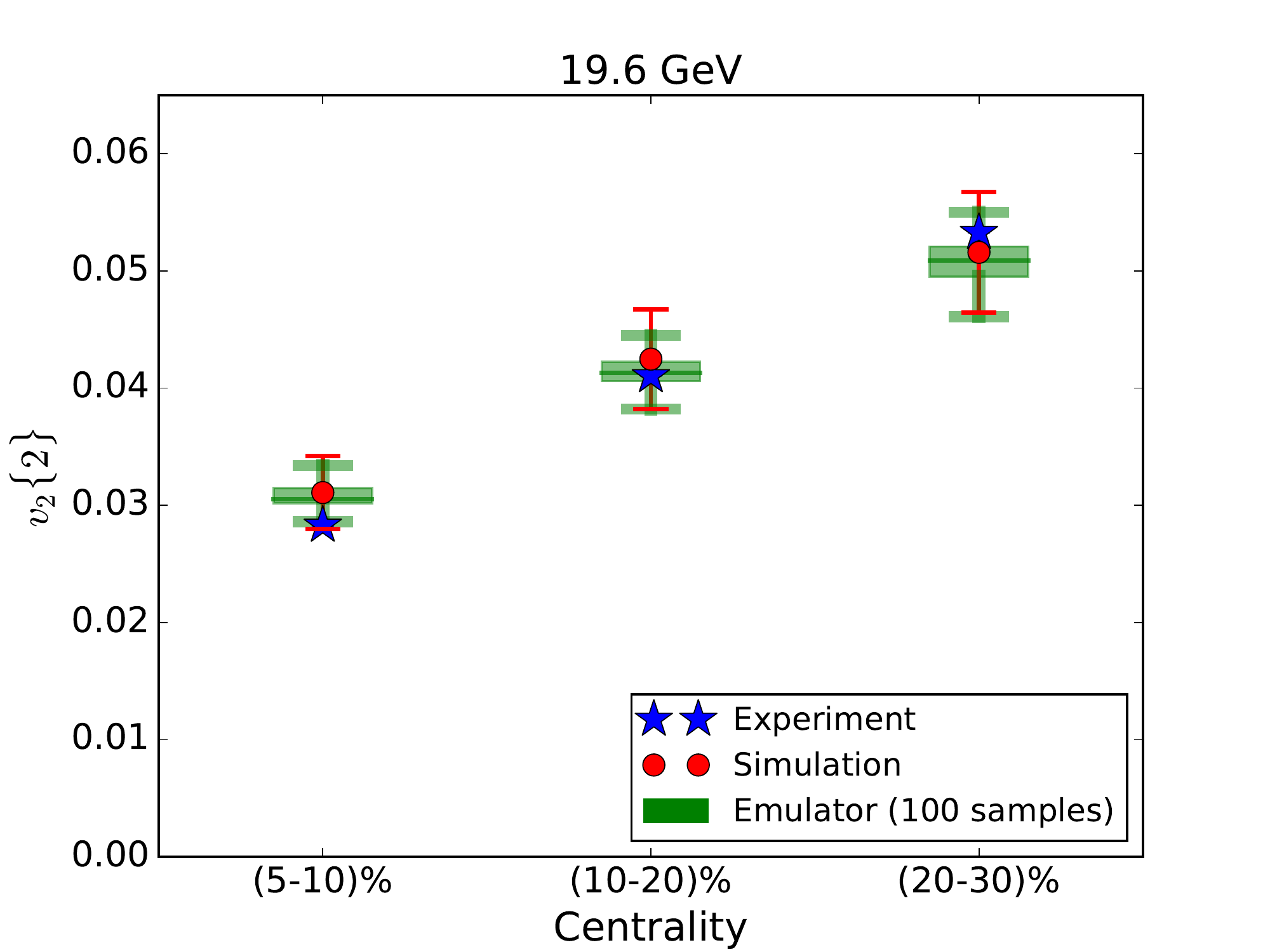}
\includegraphics[width=6.2cm]{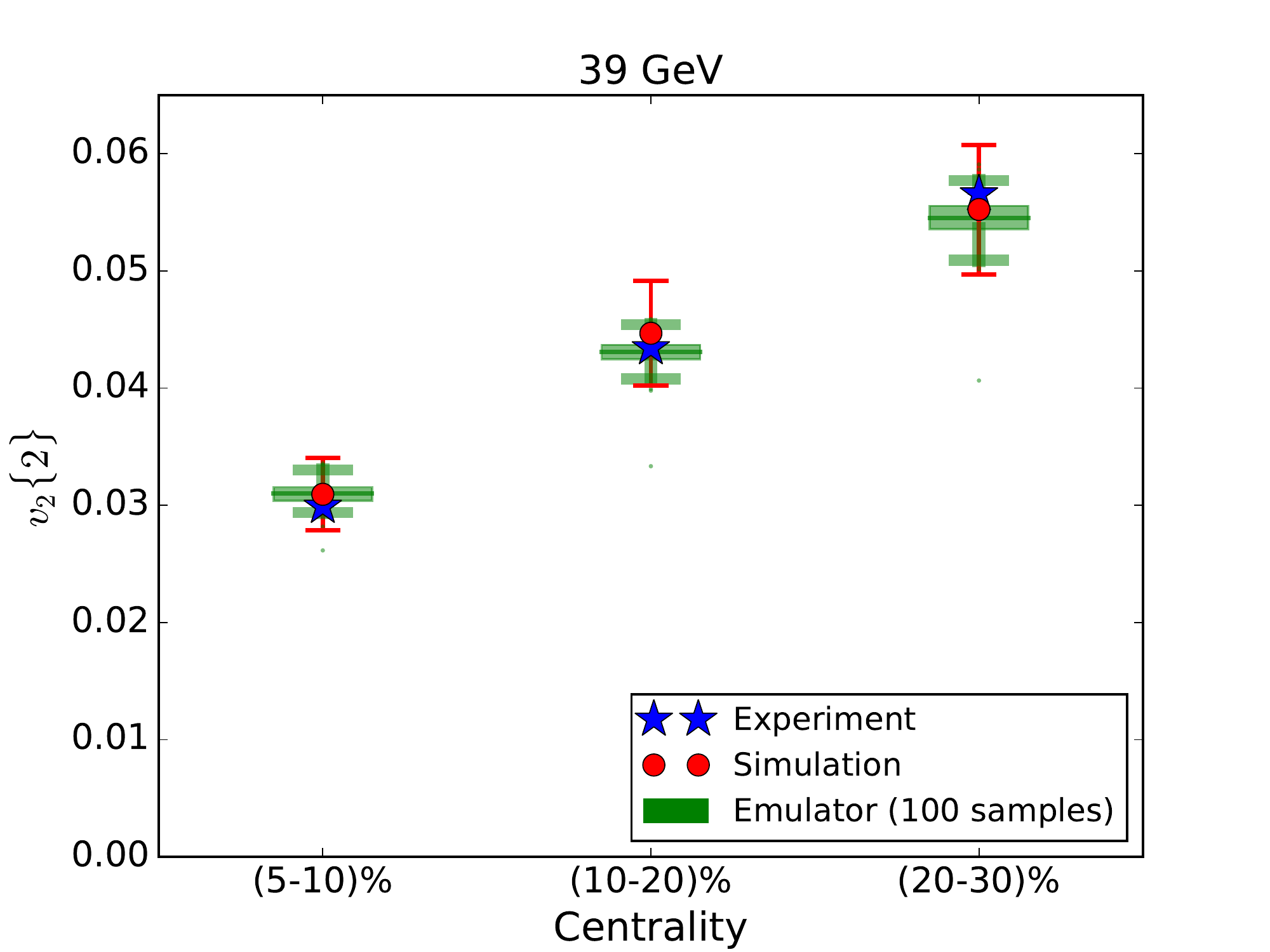}
\caption{(Color online) Analysis verification for charged particle elliptic flow $v_2\{2\}$ at $\sqrt{s_{NN}}=19.6$ and $39$ GeV.
Red dots: Simulation result using posterior median values with 10\% relative error bars.
Green box-whisker plots: Emulator predictions drawn randomly from posterior distribution.
Blue stars: STAR data \cite{Adamczyk:2012ku}.}
\label{fig:emuexp}
\end{figure}

To verify the results, 100 random parameter combinations were drawn from the posterior
and the model output for these combinations were predicted by the Gaussian process emulator.
In addition, full model simulations were run with the median values.
Example results for charged particle elliptic flow $v_2\{2\}$ at 19.6 GeV and 39 GeV are shown in Fig.~\ref{fig:emuexp}.
Emulator predictions are shown as box-and-whisker plots, where the whiskers represent the smallest and the largest 25\% of
the prediction values, and the box covers the middle 50\%.
The results demonstrate the quality of both the calibration (agreement with the experimental values is good)
and the emulation (the simulation points are close to the median of the GP predictions).

\begin{figure}
\centering
\includegraphics[width=6.2cm]{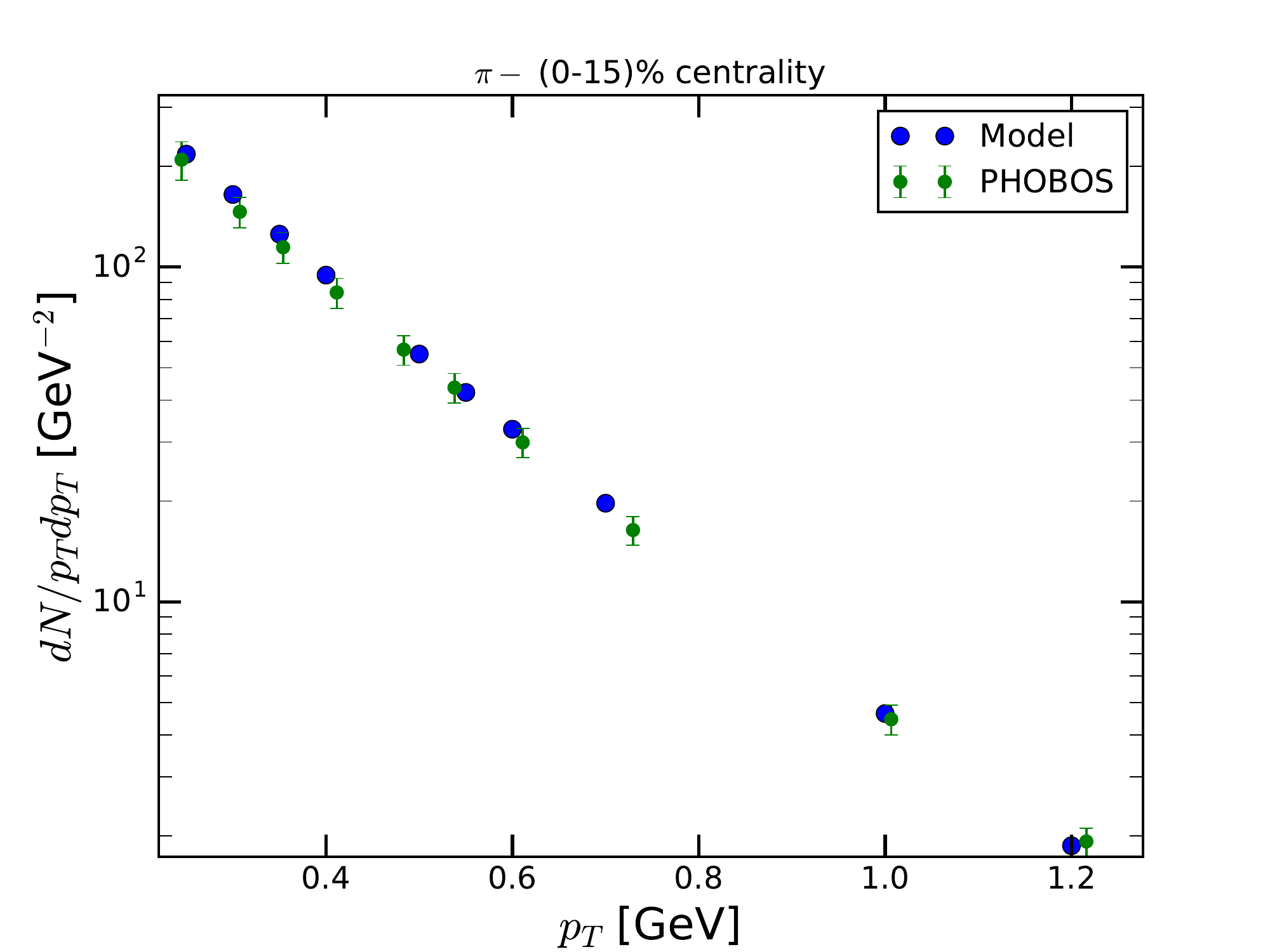}
\includegraphics[width=6.2cm]{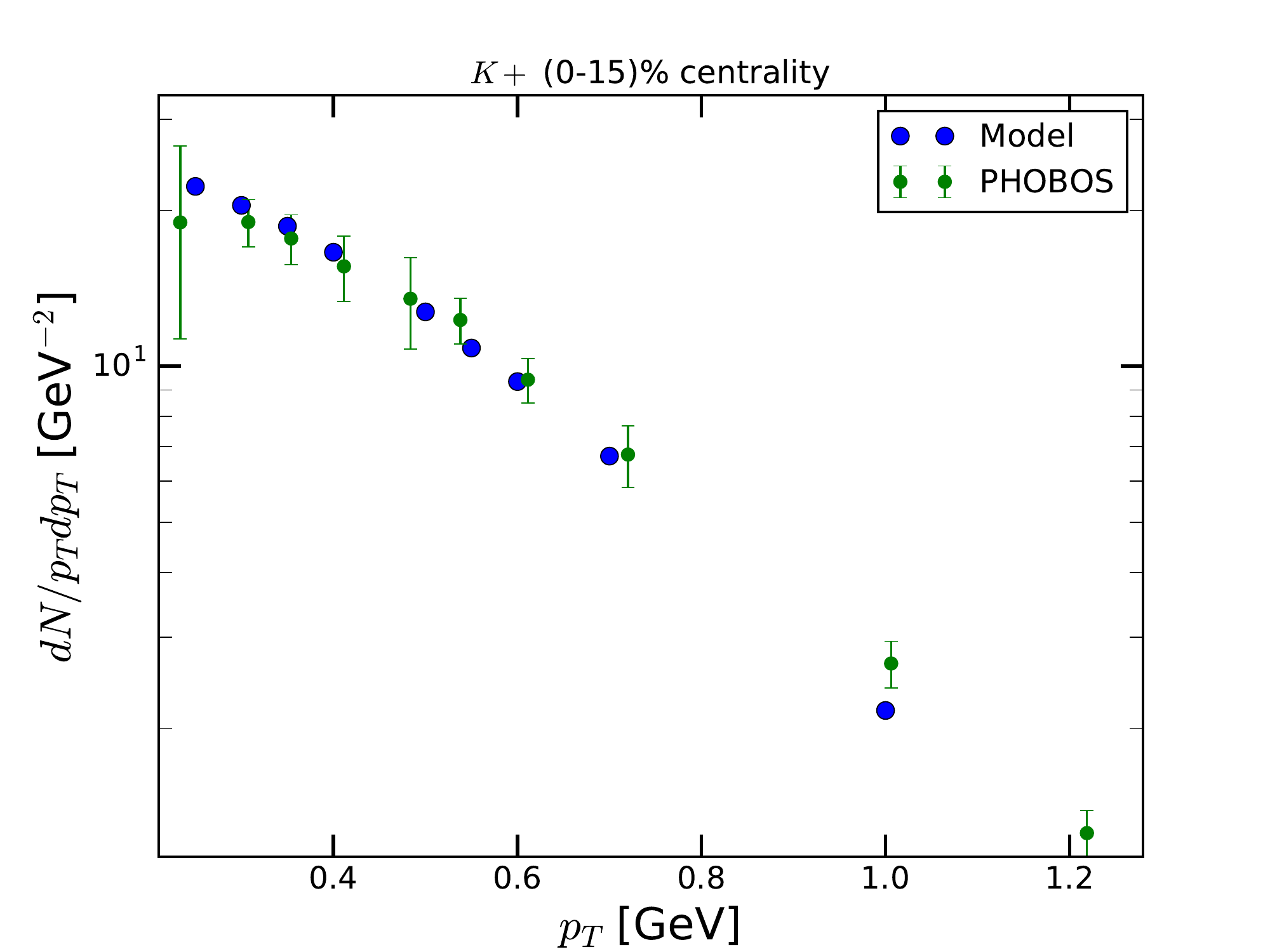}
\caption{(Color online) Transverse momentum distribution of $\pi^-$ and $K^+$ at $\sqrt{s_{NN}}=62.4$ GeV for (0-15)\% centrality.
PHOBOS data from \cite{Back:2006tt}.}
\label{fig:dndpt62}
\end{figure}

Transverse momentum distributions for $\pi^-$ and $K^+$ at $\sqrt{s_{NN}}=62.4$ GeV are shown in Fig.~\ref{fig:dndpt62}.
It should be noted that $dN/dp_T$ data were not explicitly part of the statistical analysis,
but $K/\pi$ ratio and mean $p_T$ were used instead.
It was already argued in Ref.~\cite{Novak:2013bqa} that knowledge of particle yields and mean $p_T$ should be enough to completely describe the $p_T$ spectra.
The very good agreement between model results and data seen in Fig.~\ref{fig:dndpt62} supports this argument.

\begin{figure}
\centering
\includegraphics[width=6cm]{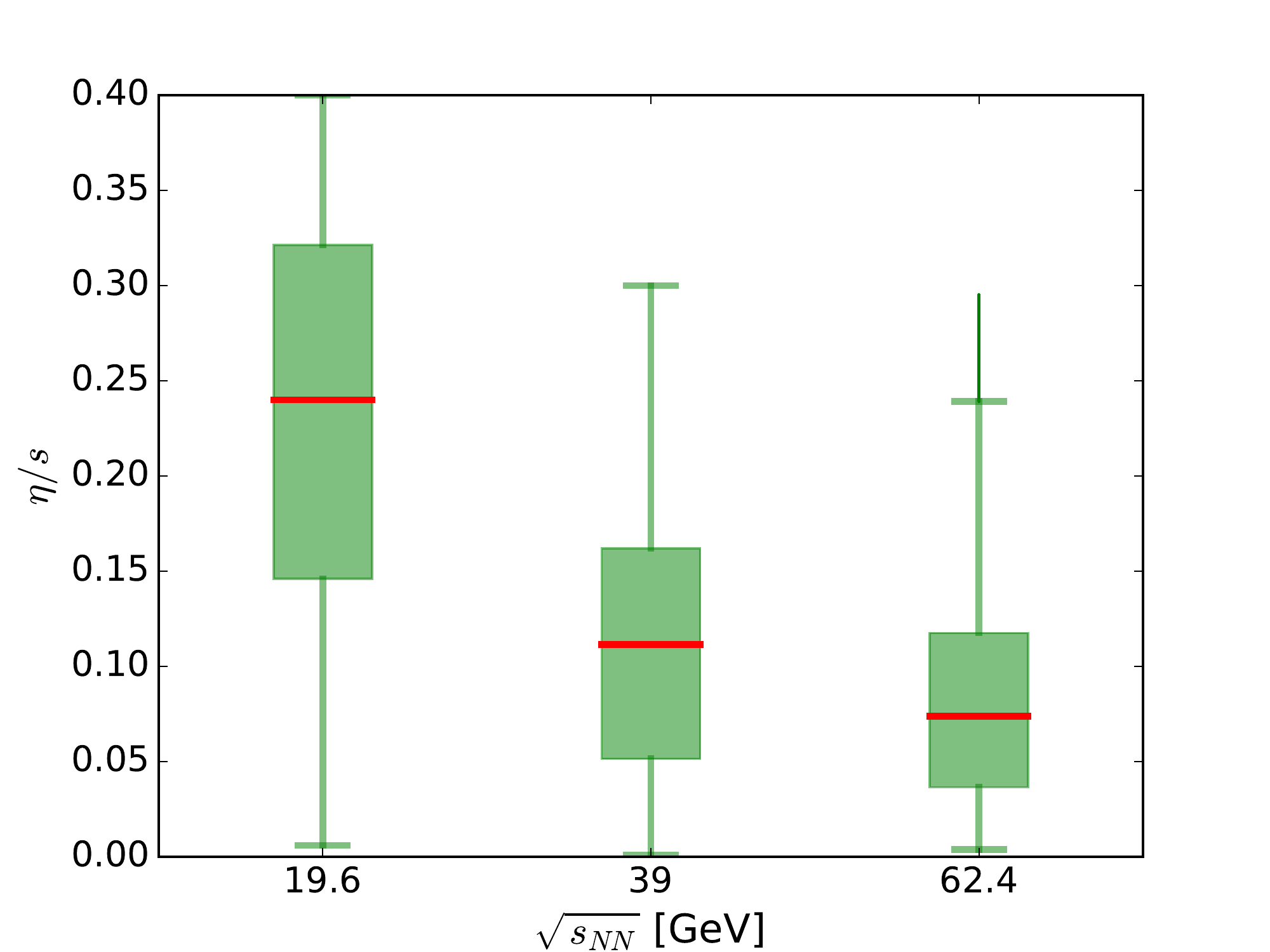}
\includegraphics[width=6cm]{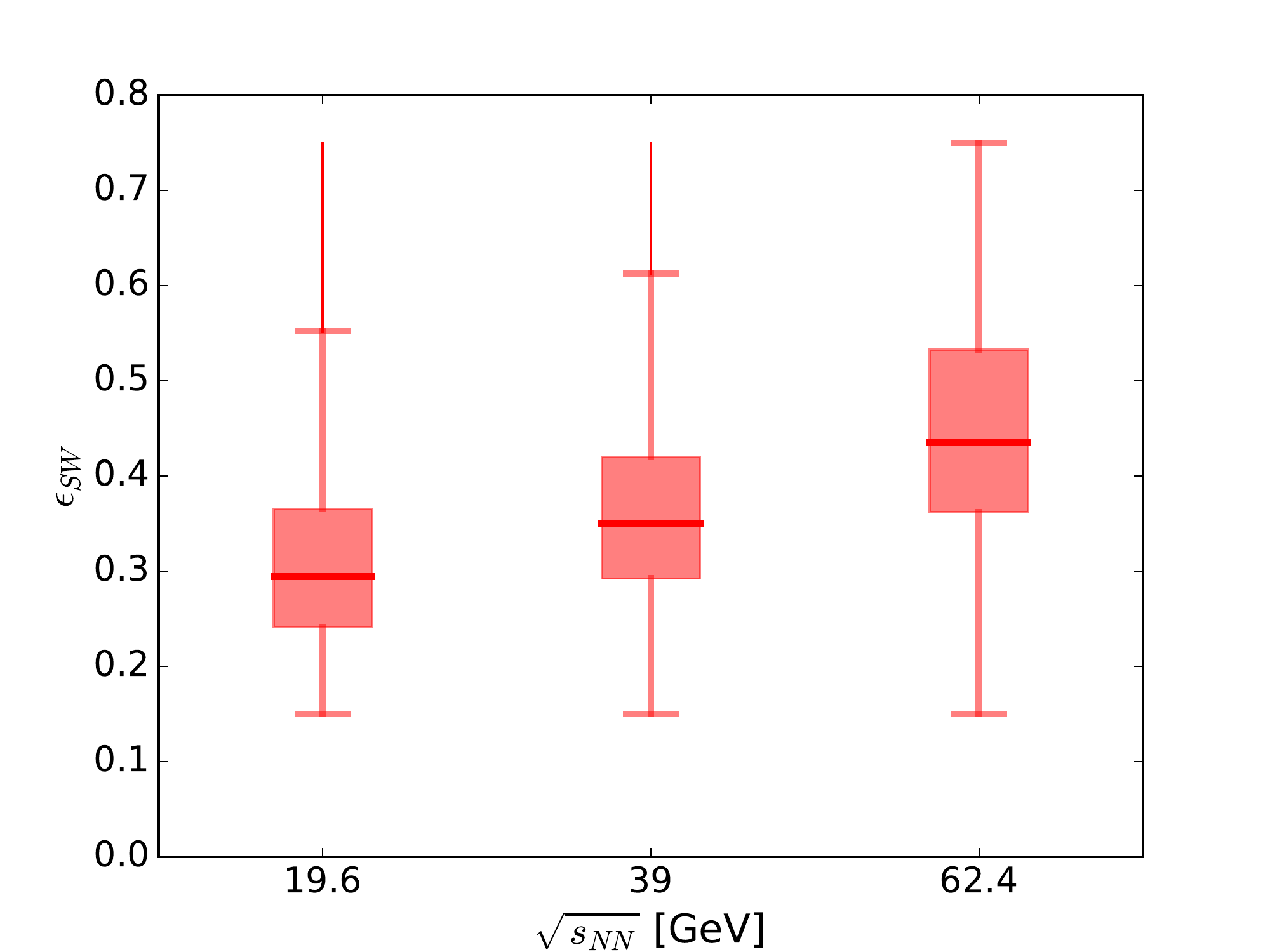}
\caption{(Color online) Box-whisker plot of the collision energy dependence of the shear viscosity over entropy density ratio $\eta/s$ and the switching energy density $\epsilon_{SW}$.}
\label{fig:bes_boxplots}
\end{figure}

The collision energy dependence of posterior distributions is illustrated in Fig.~\ref{fig:bes_boxplots},
again using the box-and-whisker representation.
No parameter values are fully excluded with the present uncertainties in likelihood calculation,
and so the range of the posterior values in most cases matches with the prior.
However, the median and peak values of $\eta/s$ move clearly towards lower values at higher collision energies,
confirming the findings in Ref.~\cite{Karpenko:2015xea}.

\section{Summary and Outlook}
\label{sec:summary}

The collision energy dependence of $\eta/s$ posterior distributions suggests that shear viscosity depends not only on temperature $T$
but also on baryochemical potential $\mu_B$,
and the parametrizations of the shear viscosity should include both dependencies.
Once such a parametrization is formulated,
it should be possible to find parameter values which fit the whole beam energy scan range simultaneously.

Current analysis focus was on the properties of bulk QCD matter and utilized
only RHIC-BES data on soft hadrons.
To improve the constraints on model parameters, data from more beam energies
and asymmetric collision systems such as p+Pb needs to be included.
In order to reduce theoretical uncertainties,
our understanding of the initial state needs to be improved
and a realistic EoS that has the proper $\mu_B=0$ limit must be included.

It is also important to note that the statistical analysis is model agnostic,
allowing us in the future to perform quantitative comparisons between multiple models
and verify/falsify different conceptual approaches within one common framework.

\section{Acknowledgements}
We thank Iurii Karpenko for providing the 3+1D viscous hydrodynamics hybrid model code.
This work has been supported by NSF grant no.~PHY-0941373 and by DOE grant no.~DE-FG02-05ER41367.
CPU time was provided by the Open Science Grid, supported by DOE and NSF.

\end{document}